\documentstyle[11pt,aaspp4,tighten]{article}

\slugcomment{CNU-AST-96}

\begin{document}

\font\twelvei = cmmi10 scaled\magstep1 
       \font\teni = cmmi10 \font\seveni = cmmi7
\font\mbf = cmmib10 scaled\magstep1
       \font\mbfs = cmmib10 \font\mbfss = cmmib10 scaled 833
\font\msybf = cmbsy10 scaled\magstep1
       \font\msybfs = cmbsy10 \font\msybfss = cmbsy10 scaled 833
\textfont1 = \twelvei
       \scriptfont1 = \twelvei \scriptscriptfont1 = \teni
       \def\mit{\fam1 }
\textfont9 = \mbf
       \scriptfont9 = \mbfs \scriptscriptfont9 = \mbfss
       \def\bmit{\fam9 }
\textfont10 = \msybf
       \scriptfont10 = \msybfs \scriptscriptfont10 = \msybfss
       \def\bmsy{\fam10 }

\def\etal{{\it et al.~}}
\def\eg{{\it e.g.}}
\def\ie{{\it i.e.}}
\def\lsim{\raise0.3ex\hbox{$<$}\kern-0.75em{\lower0.65ex\hbox{$\sim$}}} 
\def\gsim{\raise0.3ex\hbox{$>$}\kern-0.75em{\lower0.65ex\hbox{$\sim$}}} 
 
\title{Zero Energy Rotating Accretion Flows near a Black Hole
\footnote[1]{Submitted to the Astrophysical Journal}}

\author{Dongsu Ryu}
\affil{Dept. of Astronomy \& Space Sci., Chungnam National Univ., 
   Daejeon 305-764, Korea;\\
   e-mail: ryu@sirius.chungnam.ac.kr}
\author{Sandip K. Chakrabarti}
\affil{Tata Institute of Fundamental Research, Mumbai 400005, India;\\
   e-mail: chakraba@tifrc2.tifr.res.in}
\and
\author{Diego Molteni}
\affil{Department of Physics, University of Palermo, 90100 Palermo, Italy;\\
   e-mail: molteni@gifco.fisica.unipa.it}

\vskip 1cm
\begin{abstract}

We characterize the nature of thin, axisymmetric, inviscid,
accretion flows of cold adiabatic gas with zero specific energy
in the vicinity of a black hole by the specific angular momentum. 
Using two-dimensional hydrodynamic simulations in 
cylindrical geometry, we present various regimes in which 
the accretion flows behave distinctly differently. 
When the flow has a small angular momentum
$(\lambda\lsim\lambda_b)$, most of the material is accreted into 
the black hole forming a quasi-spherical flow or a simple disk-like
structure around it.
When the flow has a large angular momentum (typically,
larger than the marginally bound value, $\lambda\gsim\lambda_{mb}$),
almost no accretion into the black hole occurs.
Instead, the flow produces a stable standing shock with one or more
vortices behind it and is deflected away
at the shock as a conical outgoing wind of higher entropy.
If the flow has an angular momentum somewhat smaller than $\lambda_{mb}$
$(\lambda_{u}\lsim\lambda\lsim\lambda_{mb})$, a fraction (typically,
$5-10$\%) of the
incoming material is accreted into the black hole, but the the flow
structure formed is similar to that as for $\lambda\gsim\lambda_{mb}$.
Some of the deflected material is accreted back into
the black hole, while the rest is blown away as an outgoing wind.
These two cases with $\lambda\gsim\lambda_u$ correspond those studied
in the previous works by Molteni, Lanzafame, \& Chakrabarti (1994) and
Ryu \etal (1995).
However, the flow with an angular momentum close to the marginally stable
value $(\lambda_{ms})$ is found to be unstable.
More specifically, if $\lambda_b\lsim\lambda\sim\lambda_{ms}\lsim\lambda_u$,
the flow displays a distinct periodicity in the sense
that the inner part of the disk is built and destroyed regularly.
The period is roughly equal to $4-6\times10^3R_g/c$ depending on the 
angular momentum of the flow. 
In this case, the internal energy of the flow around the black hole
becomes maximum when the structure with the accretion shock and
vortices is fully developed.
But the mass accretion rate into the black hole reaches a maximum value
when the structure collapses.
Averaged over periods, more than a half of the incoming material is
accreted into the black hole.
We suggest the physical origin of these separate regimes from a global
perspective.
Then, we discuss the possible relevance of the instability
work on the Quasi-Periodic Oscillations (QPOs).
 
\end{abstract}

\keywords{Accretion, Accretion Disks - Black Hole Physics - Hydrodynamics -
Shock Waves}

 
\section{Introduction}

Accretion onto compact objects is an important ingredient in many
astrophysical systems involving mass transfer from one object to
another such as in binary systems or from the surrounding medium to
the center object such as in active galactic nuclei,
and hence has been the topic of many studies, both analytic and numerical.
Bondi (1952) first considered the accretion of adiabatic gas into a
gravitating point mass, $M_{bh}$, with the assumptions of non-relativistic
(but Newtonian) treatment and spherical symmetry, and found a shock-free
solution with a steady state accretion rate given by
\begin{equation}
\dot M_{acc} = 4\pi\alpha{\left(GM_{bh}\right)^2\rho\over c_s^3}_,
\end{equation}
where $\rho$ and $c_s$ are the density and sound speed of the gas at infinity.
Here, $\alpha$ is a numerical constant of order unity.

Studies of more realistic accretion of gas with a finite amount of
angular momentum, however, indicated the existence of
the variety and complexity of structures forming around the central object.
Hawley and collaborators (Hawley, Smarr, \& Wilson 1984a, 1984b;
Hawley 1986; Hawley \& Smarr 1986) demonstrated through
numerical simulations that, as accretion flow approaches the central
object, it encounters an increasing centrifugal force and material is
reflected away before reaching the black hole horizon if its angular
momentum is sufficient enough. 
In their simulations, the accretion shock where the material is
deflected is non-steady and travels outwards.

Full behavior of the accretion flows with a finite amount of
angular momentum (both for optically thick and optically thin flows) were
understood later through analytical work of basically one-dimensional flows
which are in vertical equilibrium (Chakrabarti 1989, 1990).
There it was shown that two parameters, namely, specific energy and
specific angular momentum are required to define the critical accretion rates.
It was also observed that there is a large region of the parameter space
spanned by the specific energy and the specific angular momentum of the
flows, where the accretion shock could be perfectly stable.
These results were verified later with one-dimensional and two-dimensional
numerical simulations based on the Smoothed Particle Hydrodynamics (SPH)
and Total Variation Diminishing (TVD) methods (Chakrabarti \& Molteni
1993; Molteni, Lanzafame, \& Chakrabarti, 1994; Ryu \etal 1995, hereafter
Paper I; Molteni, Ryu, \& Chakrabarti 1996). 
These works showed that weakly viscous quasi-spherical accretion 
(\eg, optically thick accretion disks of Paczy\'nski \& Wiita [1980] 
or optically thin ion tori of Rees \etal [1982]) can first form a giant
thick disk, and close to the black hole another smaller thick disk
in the post-shock region (if the shock exists).
The accretion is accompanied by a conical outgoing wind depending 
on the geometry and the specific angular momentum of incoming flow.
In conformity with the analytical work, shock free flows passing only 
through the inner or outer sonic point were also found to be stable
at least in one-dimension.

Furthermore, it was shown that the accretion shock may continue to
persist in weakly or strongly viscous flows depending on viscosity
prescriptions (Chakrabarti, 1990; Chakrabarti \& Molteni, 1995) if
the angular momentum at the outer boundary is kept sub-Keplerian
(though the result is 
not certainly restricted to this outer boundary condition).
On the other hand, two-dimensional simulations of strongly viscous
Keplerian disks found no accretion shock (Eggum, Coroniti, \& Katz
1985, 1987, 1988).
Recently, Igumenshchev, Chen, \& Abramowicz (1996) and
Manmoto \etal (1996) studied the numerical simulations of
the advection dominated flows (Narayan \& Yi 1994). 
Since the basic equations solved were similar as those in
Chakrabarti \& Molteni (1995), these works also showed the re-distribution
of angular momentum (Igumenshchev, Chen, \& Abramowicz \etal 1996) and
disappearance of the shock at high viscosity (Manmoto \etal 1996) as
shown in Chakrabarti \& Molteni (1995).
This is consistent with the understanding that the advection dominated flows
constitute a special case of viscous advective disks
(see, Chakrabarti 1996).
The general behavior in the latter flows is that
the parameter space which forms standing shock waves `shrinks'
as viscosity is increased, and beyond a critical viscosity
the shock disappears and a Keplerian disk forms
passing through the inner sonic point.

The structure with an accretion shock around the black hole could
be unstable in some situations.
We demonstrated in Paper I that the accretion flow with a large
angular momentum $(\lambda\gg\lambda_{mb})$ is unstable if the incoming
flow has a small thickness.
Molteni, Sponholz, \& Chakrabarti (1996) showed that accretion
flow can be unstable, if cooling is not negligible and its time
scale is comparable to the dynamical time scale of the post-shock flow.
These instabilities, either dynamically induced as in Paper I
or thermally induced as in Molteni, Sponholz, \& Chakrabarti (1996)
result in the oscillatory behavior of the structure formed around the
black hole causing the energy output from it to vary quasi-periodically.
These oscillating disks may be capable of producing right frequencies
and amplitude modulations for the QPO phenomenon seen in wide variety of
objects such as the low mass x-ray binaries (\eg, Van der Klis 1989;
Dotani 1992), white dwarf candidates (\eg, Mauche, 1995, 1996),
as well as in active galaxies (\eg, Halpern \& Marshall 1996).

In Paper 1, we considered the properties of thin, rotating accretion flows 
near a Newtonian point mass.
The angular momentum used was larger than the marginally bound value
($\lambda_{mb}=2R_gc$ where $R_g$ is the Schwarzschild radius).
We obtained a structure with shocks and winds consistent with the
expectations of earlier work in this angular momentum regime
(Hawley \& Smarr 1986; Molteni, Lanzafame, \& Chakrabarti 1994).
As an extension of Paper 1, in this paper we study the properties of
thin, rotating, accretion flows around {\it a black hole}.
Here we deal with transonic flows which can have an extra saddle type
sonic point just outside the horizon (Liang, \& Thompson 1980;
Chakrabarti 1989, 1990) through which the material can accrete as well. 
Especially, we focus on the cold flows (before shocking, if shock exists)
which are at rest at infinity, \ie, cold flows with negligible net energy.
Then, whether a disk will be accreting at all or whether a disk will
include a stationary/non-stationary shock will then depend only on a
single parameter, namely, the specific angular momentum of the flows.
Here, we demonstrate that the accretion flow with angular momentum somewhat
larger or smaller than the marginally stable value ($\lambda_{ms}=
\sqrt{(3/2)^3}R_gc\equiv1.837R_gc$ for a Schwarzschild black hole) establish
a stable structure around the black hole, while the flows with angular
momentum close to $\lambda_{ms}$ show an unstable behavior exhibiting
periodic construction and destruction of the inner accretion disk.
We anticipate that this regime of `unstable flows' would be very important
from astrophysical points of view, as radiation emitted from this region
could be periodically modulated producing a QPO behavior, particularly
that the frequency and amplitude of the modulation {\it are} similar to
the corresponding observed values.

The plan of the paper is the following:
In \S2, we describe the problem by defining the normalization units and
identifying the free parameters.
We then demonstrate the existence of different angular momentum regimes,
where accretion flows are expected to behave differently, using an
analytical consideration.
In \S3, we describe briefly the numerical scheme and present the results
of numerical simulations for accretion flows with different angular
momentum.
Then, we suggest the physical origin of the behavior of the accretion flows
seen in the numerical calculations.
We show that the region of the parameter space where such unstable
behavior is seen actually corresponds to the region where two saddle
sonic points exist but the stationary shock condition is {\it not}
satisfied (Chakrabarti, 1989). 
Finally, in \S4, we discuss astrophysical implication of this
work and make concluding remarks.

\section{Formulation of the Problem}

We consider the hydrodynamics of axisymmetric accretion flows
under the influence of a black hole with mass $M_{bh}$ located at the center.
We choose cylindrical coordinates $(r,\theta, z)$.
We assume that the gravitational field of the black hole can be described
in terms of the potential introduced by Paczy\'nski \& Wiita (1980),
\begin{equation}
\phi(r,z) = -{GM_{bh}\over R-R_g}_,
\end{equation}
where $R=\sqrt{r^2+z^2}$ and the Schwarzschild radius is given
by $R_g=2GM_{bh}/c^2$.
The gas mass in the accreting flow is assumed to be much smaller than
the black hole mass, so the self-gravity of the gas is negligible
compared to the gravity of the black hole.
Also the accreting matter is assumed to be adiabatic gas without
cooling and dissipation, and is described with a polytropic equation of
state, $P=K \rho^{\gamma}$, where $\gamma$ is the adiabatic index which
is considered to be constant with $4/3$ throughout the flow.
$K$ is related to the specific entropy of the flow, $s$, and varies
only at shocks (if present).
Since we consider only weak viscosity limit, the specific angular momentum of 
the accretion flow, $\lambda=rv_{\theta}$, is  assumed to be conserved.
With these settings, the problem has three intrinsic quantities,
the black hole mass $M_{bh}$, the speed of light $c$, and the Schwarzschild
radius $R_g$.
In what follows, we use them to be the units of mass, velocity, and
length respectively.

With this normalization, the condition of the incoming gas can be
characterized by a minimum of the following four parameters at a distant
point from the black hole (or, the outer boundary of the domain of
numerical calculations, $r_b$, see the next section):
1) the radial velocity, $v_r$, 2) the angular momentum, $\lambda$, 3) the
sound speed, $a$, and 4) the geometry (or more specifically, the thickness,
$h_{in}$ of the inflow), if the the complication of the vertical density
structure is ignored and the vertical velocity, $v_z$, is negligible.

The above number of the free parameters is still too large to explore
with numerical simulations, and we reduce it with the following further
assumptions:
First, we consider the gas which are at rest with negligible pressure at
infinity and start to accrete into the black hole.
So we assume that the flows have zero total energy at the outer boundary.
This assumption implies that for a given specific angular momentum,
$v_r$ and $a$ are not independent but related by the following condition,
\begin{equation}
{\cal E} = {v_r^2\over2}+{a^2\over\gamma-1}
+ {\lambda^2\over2r^2}+\phi=0,
\end{equation}
with negligible $v_z$.
Second, we consider the incoming gas at the outer boundary which is cool
and hence supersonic.
We assume a fixed Mach number $M=v_r/a$ which is large $(M\gg1)$ at
the outer boundary.
Finally, we consider the inflow at the outer boundary which has a small
thickness or a small arc angle $\theta=\arctan(h_{in}/r_b)\ll1$.
With these assumptions we can fix three parameters, $v_r$, $a$, and
$h$, and we are effectively left with a single free parameter,
the specific angular momentum $\lambda$, using which we
classify the properties of accretion flows.

In order to understand the role of angular momentum, we note that
whereas in a Newtonian geometry a particle with any non-zero
angular momentum faces infinite potential barrier close to the 
symmetry axis, around a black hole the particle `starts feeling'
the barrier only if $\lambda>\lambda_{ms}$, the marginally stable
angular momentum.
This is the {\it minimum} angular momentum that is required by a
rotating particle to balance the gravity of the black hole anywhere
outside it.
The vanishing of the net force (the centrifugal plus gravitational forces),
\begin{equation}
f = {\lambda^2\over r^3}-{1\over2(r-1)^2}_,
\end{equation}
defines the Keplerian angular momentum, $\lambda_{Kep}$, at a distance
$r=r_{eq}$. 
For an accretion flow with $\lambda>\lambda_{ms}$, there are two $r_{eq}$'s.
Between them, the centrifugal force dominates over the gravitational
force, so the total force is directed outwards.
Outside them, the gravitational force is larger, so the total force is
directed inwards.
If a flow approaches from infinity, it feels the inward force at first
and then feels the outward force. Hence, the flow can 
form a stable disk around the black hole.
But if the accretion flow has $\lambda<\lambda_{ms}$, the gravitational
force always dominates over the centrifugal force, so the flow always feels
the inward attraction.
This flow, thus, can not form a stable disk, but accretes directly into
the black hole in a quasi-spherical manner.

The marginally bound angular momentum, $\lambda_{mb}$, on the other 
hand, defines the highest possible angular momentum with which a particle
at rest at infinity can be accreted into a black hole (\eg, Shapiro \&
Teukolsky 1983).
Or, inversely, it is the smallest possible angular momentum for a
rotating particle to bounce back to infinity.
The location of bouncing back as a function of $r$ from the black hole is
commonly known as the funnel wall and can be found from the vanishing of
the energy in Eq.~(3) (\eg Hawley \& Smarr 1986).
Along the equatorial plane, it is given by
\begin{equation}
{\cal E} = {\lambda^2\over2r^2} - {1\over2(r-1)} = 0.
\end{equation}
We denote the radius of the funnel wall along the equatorial plane
to be $r_f$.
If $\lambda > \lambda_{mb}$, flows are required to be {\it hot}
(${\cal E}>0$) in order to accrete onto a black hole.

In Fig.~1, we show $r_{eq}$ and $r_{f}$ in long dashed and short dashed
curves and also note the limiting angular momenta.
Since we are interested in the transonic properties of accretion flows,
we also show with dotted curve the locations of the sonic points
($r_{sonic}$) of zero energy flows.
It has been calculated using the standard analysis of one-dimensional flows
(Chakrabarti 1989, 1990; Chakrabarti \& Molteni 1993).
With $\lambda<\lambda_b(=1.782)$, accretion flows have no sonic point
at a finite distance.
With $\lambda_b\leq \lambda \leq \lambda_{mb}$, they have a saddle-type
inner sonic point, $r_s$, (dotted curve with negative slope) as well as
an unphysical center-type sonic point, $r_c$, (dotted curve with
positive slope).
With $\lambda>\lambda_{mb}$, they have only the unphysical sonic point.

Furthermore, we can classify the angular momentum space into four zones
with three boundaries at $\lambda_b=1.782$, $\lambda_u=1.854$, and
$\lambda_{mb}=2$, respectively.
For $\lambda<\lambda_b$ (Zone I), with no sonic point at a finite distance,
only supersonic flows can accrete onto a black hole without accompanying
a shock.
In the region $\lambda_u<\lambda<\lambda_{mb}$ (Zone III), the
Rankine-Hugoniot condition for shock formation is satisfied at $r>r_s$
if the inflow is supersonic (Chakrabarti 1989).
Thus, supersonic flows can settle into a stable solution which includes
a shock.
However, in the region $\lambda_b<\lambda<\lambda_{u}$ (Zone II),
the Rankine-Hugoniot condition is not satisfied.
Therefore, supersonic flows cannot settle into a stable solution.
For $\lambda>\lambda_{mb}$ (Zone IV), the inner sonic point does not exist.
Thus, although a standing shock may form if the inflow is supersonic,
matter cannot accrete onto a black hole but must be diverted as a wind.
Although the above classification is based on  the analysis for
one-dimensional flows, we find through numerical simulations that these
subdivisions may be robust even in two-dimensional flows.
And what is more, we discover that the flows in Zone II exhibit a periodic
behavior, virtually building and destroying the whole disk structure while
constantly  trying to form a standing shock wave, but never succeeding to
do so.

The classification described above is applicable to the flows of zero net
energy and the adiabatic index of $4/3$.
It can be extended to more general flows with non-zero energy and any
adiabatic index.
Thus, for instance, the unstable periodic behavior will be present in the
region of the parameter space where shock condition is not satisfied,
though the flow has two saddle type sonic points (Fig.~4 of
Chakrabarti 1989).
Details of the general classification with relevant numerical simulations
will be presented elsewhere (Chakrabarti, Ryu, \& Molteni 1996).

\section{Results of Numerical Calculations}

\subsection{Numerical Method and Setup}

The numerical calculations have been done with a hydrodynamic code
based on the TVD scheme, originally developed by Harten (1983).
It is an explicit, second order accurate scheme which is designed to
solve a hyperbolic system of conservation equations, like the system
of hydrodynamic equations.
It is a nonlinear scheme obtained by first modifying the flux function
and then applying a non-oscillatory first order accurate scheme to
get a resulting second order accuracy.
The key merit of this scheme is to achieve the high resolution of a
second order accuracy while preserving the robustness of a non-oscillatory
first order scheme.

Harten (1983) described the application of his scheme to the set
of one-dimensional hydrodynamic equations in Cartesian geometry
and presented some tests including shock capturing.
The version of the TVD code used in this paper is the two-dimensional
one in cylindrical geometry which was described in details in Paper I,
except the form of the gravitational force.
Here, we use the force which is derived from the modified potential
in Eq.~(2).
In Molteni, Ryu, \& Chakrabarti (1996), the code was tested for the
problem of one-dimensional and two-dimensional accretion flows by
comparing the solutions from the code with the analytic solutions and the
numerical solutions from the SPH code and demonstrated to be sufficiently
accurate for the numerical study of accretion flows.

The equations solved numerically with the TVD code are given in
Molteni, Ryu, \& Chakrabarti (1996).
Note that, in the case of an axisymmetric flow without viscosity,
the equation for azimuthal momentum states simply the conservation of
specific angular momentum,
\begin{equation}
d\lambda/dt=0.
\end{equation}
So it can be decoupled from the rest of the hydrodynamic equations,
if the flow has a uniform angular momentum.
But in the problem considered in this paper, the background material
(of relatively low density) and the accreting material may have 
different angular momenta and be mixed in some cells resulting in a
variation in specific angular momentum.
So the code used in this paper solves the whole set of the hydrodynamic
equations including that for azimuthal momentum.
As the results, however, the angular momentum of the accreting material is
not conserved exactly.
The error in the angular momentum conservation is typically less than
$\sim2\%$ in the material approaching the black hole, but could be up to
$\sim5\%$ in the material bounced back as a wind.
For a discussion on the possible effects of this error,
see Molteni, Ryu, \& Chakrabarti (1996).

The calculations have been done in a setting similar to that used in
Paper I.
The computational box occupies one quadrant of the $r-z$ plane
with $0\leq r\leq50$ and $0\leq z\leq50$.
The incoming gas enters the box through the outer boundary located at
$r_b=50$.
We have chosen the adiabatic index $\gamma=4/3$, since it is appropriate
for a wide range of astrophysical circumstances, such as for relativistic
flows, optically thick radiation dominated disks, and optically thin flows
with cooling effects (such as the Comptonization of external soft photons).
We have chosen the density of the incoming gas $\rho_{in}=1$ for convenience,
since, in the absence of self-gravity and cooling, the density is scaled out,
rendering the simulation results to be valid for any accretion rate. 

We have considered accretion flows with angular momentum
$1.5\leq\lambda\leq2.25$ covering all the four zones in Fig.~1.
The incoming flow at the outer boundary is directed toward the symmetry
axis along the negative $r$-axis with $v_z=0$.
It has a large Mach number, $M=10$, at the outer boundary.
This setting with the assumption of zero total energy in Eq.~(3) fixes the
values of $v_r$ and $a$ at the outer boundary for the flows with given
angular momentum.
The incoming flow has been further assumed to be thin with a uniform vertical
density.
So at the outer boundary, $r_b=50 R_g$, we inject matter for
$0<\theta<\arctan(1/10)$.
This gives the half-thickness of the incoming flow as $h_{in}=5 R_g$.

In order to mimic the horizon of the black hole at the Schwarzschild
radius, we have placed an absorbing inner boundary at $R=1.5R_g$,
inside which all the material is completely absorbed into the black hole.
For the background material, we have used a stationary gas with
density $\rho_{bg}=10^{-6}$ and sound speed (or, temperature) same as
that of the incoming material.
Hence, the incoming material has pressure $10^6$ times larger than that
of the background material.
So once the incoming flow enters through the outer boundary, it initially
expands vertically.

Table 1 summarizes the values of parameters used for the numerical
calculations along with the identification of the model number.
The duration of each model calculation is indicated as $t_{end}$ in units
of $R_g/c$.
Except for Model M9, all the calculations have been done with
$256\times256$ cells, so each grid has the size of $0.195$ in the
unit of the Schwarzschild radius.
The calculation for Model M9 has been done with $128\times128$ cells to
study the convergence of our calculations. 

\subsection{Numerical Solutions}

Fig.~2 shows the temporal evolution of the position of the accretion shock
along the equatorial plane with $z=0$, $r_{sh}$, the mass accretion rate
into the black hole, ${\dot M}_{acc}$, and the mean density, ${\bar\rho}$,
and the mean thermal energy, ${\bar e}_{th}$, of disk matter inside the
computational domain.
Here, ${\dot M}_{in}$ is the mass inflow rate at the outer boundary.
The shock position along the equatorial plan has been calculated in the
region with $r\geq3.2$.
Just for the sake of drawing the figure, if there is no shock or a shock
in $r<3.2$, the shock position has been set to be $3.2$.
Fig.~2a is for Model M2 belonging to Zone I (left panel) and Model M5
belonging to Zone III (right panel) with $\lambda=1.75$ and $1.9$.
Fig.~2b is for Model M3 with $\lambda=1.8$ and Fig.~2c for Model M4 with
$\lambda=1.85$, both models belonging to Zone II.
The accretion flow in Model M2 shows a stable behavior with virtually all
incoming material absorbed into the black hole.
The flow in Model M5 also shows a stable behavior, but with small
aperiodic oscillations.
Several different timescales could be identified, but the most dominant
one in ${\dot M}_{acc}$ is $\sim300$ in our time unit, $R_g/c$.
But, due to a small amplitude, these oscillations may not have any
observational significance.
In this model, a small fraction $(\sim10-15\%)$ of the incoming material
is absorbed into the black hole.
This is similar to the result of a thick accretion disk simulation
(Molteni, Lanzafame, \& Chakrabarti 1994). 
However, the accretion flows in Models M3 and M4 show a distinctively
periodic behavior, during which the mass accretion rate into the black
hole, the disk mass, and the disk thermal energy vary dramatically.
The time period of variation is $4-6\times10^3$ in our time unit.

Fig.~3 shows the density contours and the velocity vectors at
$t=2\times10^4$ in Model M2 (left panel) and Model M5 (right panel).
The plots show only the region with $0\leq r\leq36$ and $0\leq z\leq36$,
although the calculations have been done in the region with $0\leq r\leq50$ 
and $0\leq z\leq50$.
The density contours are drawn with $\rho=10^k$ with
$k=-3$, $-2.9$, $-2.8$, $\ldots$, $2$ with the outermost
contour near the funnel wall being of lowest density.
For clarity, the velocity vectors are drawn at every fourth grid along 
the $r$ and $z$-directions.
The flow in Model M2 with small angular momentum is directly absorbed
into the black hole forming a simple disk-like structure around it.
The flow in Model M5 with large angular momentum forms a stable accretion
shock with one or more vortices behind it, as described in our previous
work (Molteni, Lanzafame, \& Chakrabarti 1994; Paper I).
The incoming material is deflected at the shock away from the
equatorial plane.
Some of the deflected material is accreted back into the black hole
at a higher latitude but other is blown away as a conical outgoing wind.

Fig.~4 shows the density contours and the velocity vectors at
$t=1.4\times10^4$, $1.6\times10^4$, $1.8\times10^4$, and $2\times10^4$
in Model M4.
The description on the plotted region and quantities is identical
to that of Fig.~3. 
In this case, the structure with an accretion shock and a generally
subsonic high density disk is established around the black hole.
However, the structure is not stable.
At the accretion shock, the incoming flow is deflected.
But some of the post-shock flow, which is further accelerated by the
pressure gradient behind the shock, goes through a second shock,
where the flow is deflected once more downwards.
The downward flow squeezes the incoming material, and the
accretion shock starts collapsing ($t=1.4\times 10^4$, also see, Fig.~2c).
In the process of the collapse, some of the post-shock material escapes
as wind and other is absorbed into the black hole.
But all the squeezed incoming material is absorbed into the black hole.
Thus, the mass accretion rate is larger than the mass inflow rate,
${\dot M}_{acc}/{\dot M}_{in}>1$ (`hyper-accretion stage'). 
The mass accretion rate becomes highest just after the accretion shock
collapses ($\sim 1.4{\dot M}_{in}$ in Model M3 and $\sim 1.2{\dot M}_{in}$
in Model M4).
It then starts decreasing until the shocked material is completely evacuated
and ${\dot M}_{acc}\approx{\dot M}_{in}$ is reached ($t=1.6\times 10^4$
and $t=1.8\times 10^4$, also see, Fig.~2c).
After that, the re-building of the accretion shock starts with the incoming
material bouncing back from the centrifugal barrier.
The subsonic post-shock region becomes a reservoir of material,
so the mass accretion rate is reduced dramatically.
With the accumulated material in the post-shock region a giant vortex
is formed, which in turn supports the accretion shock ($t=2\times 10^4$).
This is the time when the shock radius and the disk mass are highest but 
the mass accretion is lowest (see, Fig.~2c).
This continues until the incoming flow is squeezed enough so the accretion
shock collapsed, and the cycle continues.
Averaged over periods, more than a half of the incoming material is
accreted into the black hole. 

The collapse and rebuilding of the structure takes place due to a very
complex and non-linear interaction of the inflow with the outgoing material,
whose property is determined by the accretion shock and the post-shock
dynamics.
As a result, the growth and decay of the average density and pressure follow
exponential rules as in the charging and discharging of a capacitor.
In Model M3 with $\lambda=1.8$, the average density typically rises as
\begin{equation}
{\bar \rho}_{r} = 0.30\left[1-\exp\left(-t\over400\right)\right]+0.15,
\end{equation}
and falls as
\begin{equation}
{\bar \rho}_{f} = 0.30\exp\left(-t\over400\right)+0.15,
\end{equation}
in a single cycle.
The full cycle takes about $\tau_{1.8}\approx4\times10^3R_g/c$.
In Model M4 with $\lambda=1.85$, on the other hand, the average density
rises as
\begin{equation}
{\bar \rho}_{r} = 0.42\left[1-\exp\left(-t\over750\right)\right]+0.15,
\end{equation}
and falls as
\begin{equation}
{\bar \rho}_{f} = 0.42\exp\left(-t\over750\right)+0.15.
\end{equation}
The full cycle takes about $\tau_{1.85}\approx6\times10^3R_g/c$.
These exponential curves have been superimposed on density plots
in Fig.~2b and  Fig.~2c.
One would have naively expected these time scales to be twice (collapse
and expansion) as big as the infall time
\begin{equation}
\tau_{infall}=\int_{1}^{r_{out}} {dr\over v_r},
\end{equation}
which are $2280$ and $3160$ respectively with
\begin{equation}
v_r={1\over\left[7(r^{-1}-\lambda^2 r^{-2})^{-1/2}\right]}
\end{equation}
for a zero-energy post-shock flow of $\gamma=4/3$.
Here, $r_{out}$ is the average location of the shock at full expansion.
But in reality, it takes almost twice as much, possibly because the back
flowing matter has comparable ram pressure which slows down the collapse
and expansion significantly.

Fig.~5 shows the density contours and the velocity vectors at
$t=2\times10^4$ in Model M7 (left panel) and Model M9 (right panel).
The plots show the region with $0\leq r\leq50$ and $0\leq z\leq50$.
The description on the plotted quantities is identical to that of Fig.~3. 
The plots represent the same calculation for the flow in the boundary of
Zones III and IV but with different resolutions.
Overall they look similar to that of Model M5 in Zone III (see, Fig.~3),
although the flow has a much smaller accretion onto the black hole.
The general picture agrees with the previous works by Molteni, Lanzafame,
\& Chakrabarti (1994) and in Paper I.
Here there is a structure with an stable accretion shock with one or more
vortices behind it, and a generic outgoing wind for a wide range of angular
momentum.
Globally the solutions with two different resolutions
look similar to each other, however, while details vary.
This indicates that our calculations have converged, at least globally.

Fig.~6 shows some quantities of importance as a function of the
specific angular momentum.
The plots show the time-averaged values of the position of the accretion
shock along the equatorial plane, the mass accretion rate into the
black hole, the mean density of disk matter, and the mean thermal energy
of disk matter.
As the specific angular momentum increases, the shock moves further out
and the mass accretion rate decreases, as expected.
The disk mass continues to increase with the specific angular momentum.
This is because the shock moves out and 
there is an increasing space where the material is accumulated.
However, the disk thermal energy per unit volume reaches a maximum
at $\lambda\sim2$ (and the disk thermal energy per unit mass
reaches a maximum at $\lambda\sim1.95$), indicating that the piling up
of energy in the disk is most efficient if the flow has
$\lambda\sim\lambda_{mb}$.

\subsection{Comparison with Analytic Solutions}

It is pertinent to ask how the numerical solutions of the time-dependent
problem described in the previous subsection are compared to the steady
state analytic solutions of the similar set of equations.
If the flows are strictly one-dimensional without vertical motion,
the numerical solutions agree perfectly with the analytic ones as
shown in Molteni, Ryu, \& Chakrabarti (1996).
The flows considered in this paper are, however, two-dimensional, which
can expand away from the equator and can form winds as well.
The situation becomes more complicated, since vortices with back flows
can develop and interact with the incoming matter.

In the upper panel of Fig.~7, we show the analytical solutions
(Mach number vs.~radial distance) of one-dimensional, zero-energy flows
with Mach number $M=10$ at $r=50$ whose angular momentum ranges from
$\lambda=1.5$ to $2$.
With $\lambda=1.5$ and $1.75$ (the uppermost solid and the next short
dash-dotted curves), the flows always remain supersonic ($M>1$).
In these cases, the numerical simulations have found a stable solution
(Models M1 and M2). 
Shocks are not expected, if the flow is strictly one-dimensional.
Shocks are possible only in a thick flow in vertical equilibrium (see,
Chakrabarti 1989).
We also see a shock formation in the simulations, possibly
due to the change in flow geometry through vertical expansion.
The short dashed and the long dash-dotted curves are for the flows with
$\lambda=1.8$ and $1.85$ respectively.
Though solution branches through the inner sonic point exist, the upper
supersonic branch cannot be connected with the corresponding lower subsonic
branch through the shock condition, since the Rankine-Hugoniot
conditions are {\it not} satisfied.
Therefore, a steady solution with an accretion shock is impossible.
In the simulations (Models M3 and M4), we have seen an unstable disk structure
with an accretion shock which is constructed and destructed periodically.
In the lower panel of Fig.~7, we re-draw the analytical Mach number
distributions for $\lambda=1.85$ (solid curves) along with the numerical
ones on the equatorial plane (dashed curves) at four different phases
in Model M4 (at the same four epochs at which the panels of Fig.~4
was drawn).
It is clear that the flow constantly tries to adjust itself to follow the
analytical solution during the whole cycle.
What actually happens is that the shock is never pressure-balanced and the
excess or deficit of total pressure continuously drives the shock to one
direction or the other.
Also, in the upper panel, the steady state solutions for $\lambda=1.9$
and $1.95$ (long dashed and long-short dashed curves respectively)
are shown.
In these cases, the supersonic branch and the subsonic branch can
be joined together to form a standing shock (indicated with vertical dotted
lines at $r=13.1$ and $38.9$ respectively),
as described in Chakrabarti (1989) and Chakrabarti \& Molteni (1993).
In the simulations (Models M5 and M6), we do see a standing
accretion shock as well.
But the shock is located farther out, possibly due to the presence
of vortex pressure as discussed in Molteni, Lanzafame, \& Chakrabarti
(1994).
The lowermost solid curve which crosses $M=1$ line almost vertically is
for $\lambda=2$ and clearly does not have the inner sonic point.
As a result, although the supersonic inflow can have a strong standing
shock, matter does not accrete onto the black hole and must be deflected
as winds.

\section{Astrophysical Implications and Concluding Remarks}

Accretion flows are expected to take place with various inflow
conditions.
In this paper, we have discussed the effects of varying angular
momentum in thin cold flows with vanishing net specific energy.
We have divided the angular momentum space into four Zones according to
whether the flows allow sonic points and shocks.
For the flows in Zones I, III and IV, stable solutions have been found.
However, for the flows in Zone II, numerical simulations have yielded
an unstable solution.
Here, the flows periodically build and destroy the structure with
the accretion shock in a time scale of several infall times.
Two important properties of this Zone point to its generic nature:
(1) The range of angular momentum (from $\lambda_b=1.782$ to
$\lambda_u=1.854$ for one-dimensional flows) is around the marginally
stable value, $\lambda_{ms}=1.837$, which is believed to be very natural
for accretion flows.
(2) In the limit of weak viscosity and negligible cooling (as we are
considering here), our result is independent of the mass inflow rate.

The time scale of the periodicity of unstable flows is even more
interesting.
As noted in the previous section, it is in the range of
\begin{equation}
\tau \approx 4-6\times10^3{R_g\over c}=4-6\times10^{-2}
\left({M_{bh}\over M_{\sun}}\right){\rm s}.
\end{equation}
The modulation of amplitude is also very significant, and could be
as much as a hundred percent depending on detailed processes.
Oscillations with these characteristics have been observed in black hole
candidates and are called the QPOs.
In the QPOs from the low mass x-ray binaries, the oscillation frequency
has been found to lie typically between 5 and 60 Hz (Van der Klis 1989;
Dotani 1992; Mauche, 1995, 1996). 
Thus, compact objects with mass $M \approx 0.3-5M_\odot$ could 
generate oscillations of right frequencies due to the instability
discussed in this paper.
Similarly, the QPOs with time period of $\sim 1{\rm d}$ was reported
in Seyfert Galaxy RX J0437.5-4711 (Halpern \& Marshall 1996) which also 
has the right frequency, energy range (extreme UV) as we expect from the
emission property of the post-shock region.
If, on the other hand, the oscillations are due to the shocks of the
outflowing winds, the time period of oscillations would be shorter
(by, say, an order of magnitude), since the corresponding stable shock 
locations in winds are closer to the black hole.
However, by considering simplified physics which we have assumed here,
it may be premature to assume that the presented mechanism would explain
all the QPOs observed.
As a future work, we plan to include radiative processes and viscosity in
the accretion calculations to examine the observational consequences of the
present instability in more details.

Note that there exist other mechanisms suggested to explain the QPOs.
Especially, those for the multi-dimensional accretion flows include the
dynamically induced instability suggested in Paper I and the thermally
induced instability suggested in Molteni, Sponholz, \& Chakrabarti (1996).
The first one works if the accretion flow with $\lambda\gg\lambda_{mb}$
has a small inflow thickness, while the second one works if the flow
with non-negligible cooling has a cooling time scale comparable to the
dynamical time scale.
As a result, the second depends on the accretion rate, while the first
and the one in this paper are almost independent of the accretion rate
(although indirect dependence comes in as the region of parameter space
which does not produce shocks itself depends on cooling processes).
Particularly interesting is that the above mentioned oscillations
of Molteni, Sponholz and Chakrabarti (1996) occur
in the parameter space where a steady state analytical shock condition
{\it is satisfied} unlike the situations of Models M3 and M4 where the 
steady state shock condition is not satisfied. 
Recently, it is also speculated that the acoustic perturbation of 
accretion shock waves (Yang \& Kafatos 1995) may also cause QPO phenomena,
although the amplitude of modulation may be smaller.
It is possible that different observed QPOs may be due to one or more of
these mechanisms. 

The discussion presented here is valid for inviscid flows or flows with
a small viscosity.
We believe that a small viscosity should not change our conclusions. 
As we have already discussed, with a larger viscosity, the steady state
shock solutions disappear.
It is not clear if the same conclusions hold for the shocks in non-steady
solutions.
Since the driving force which causes the oscillations is the imbalance
between the total pressure on both sides of the shock, it is quite likely
that the oscillations will persist even in the presence of large viscosity.
In future, we plan to report the results of such a study.

\acknowledgments 

DR thanks J.~P.~Ostriker for providing initial guidance of the work.
The work by DR was supported in part by the Basic Science Research
Institute Program, Korean Ministry of Education 1995, Project
No.~BSRI-95-5408.

\clearpage

\begin{deluxetable}{ccccccc}
\footnotesize
\tablecaption{Summary of Simulations Reported}
\tablehead{\colhead{Model\tablenotemark{a}} & \colhead{Zone} &
\colhead{$\lambda$} & \colhead{resolution} & \colhead{$v_r$} &
\colhead{$t_{end}$} & \colhead{comments}} 
\startdata
M1 & I & $1.5~$ & $256\times256$ & 0.13566 & $2\times10^4$ 
& stable accretion  \nl
M2 & I & $1.75$ & $256\times256$ & 0.13453 & $2\times10^4$
& without wind \nl
\hline
M3 & II & $1.8~$ & $256\times256$ & 0.13428 & $4\times10^4$
& periodic \nl
M4 & II & $1.85$ & $256\times256$ & 0.13402 & $4\times10^4$
& behavior\nl
\hline
M5 & III & $1.9~$ & $256\times256$ & 0.13376 & $2\times10^4$
& stable structure with \nl
M6 & III & $1.95$ & $256\times256$ & 0.13348 & $2\times10^4$
& accretion and wind \nl
\hline
M7 & IV & $2.0~$ & $256\times256$ & 0.13320 & $2\times10^4$
& stable structure with \nl
M8 & IV & $2.25$ & $256\times256$ & 0.13169 & $2\times10^4$
& wind but without \nl
M9 & IV & $2.0~$ & $128\times128$ & 0.13320 & $6\times10^4$
& significant accretion \nl
\enddata

\tablenotetext{a}{All models have used $\gamma=4/3$ and ${\cal E}=0$,
$M=10$ and $h_{in}=\arctan(1/10)$ at the boundary, $r_b=50$.}
 
\end{deluxetable}

\clearpage

\begin{center}
{\bf FIGURE CAPTIONS}
\end{center}
\begin{description}

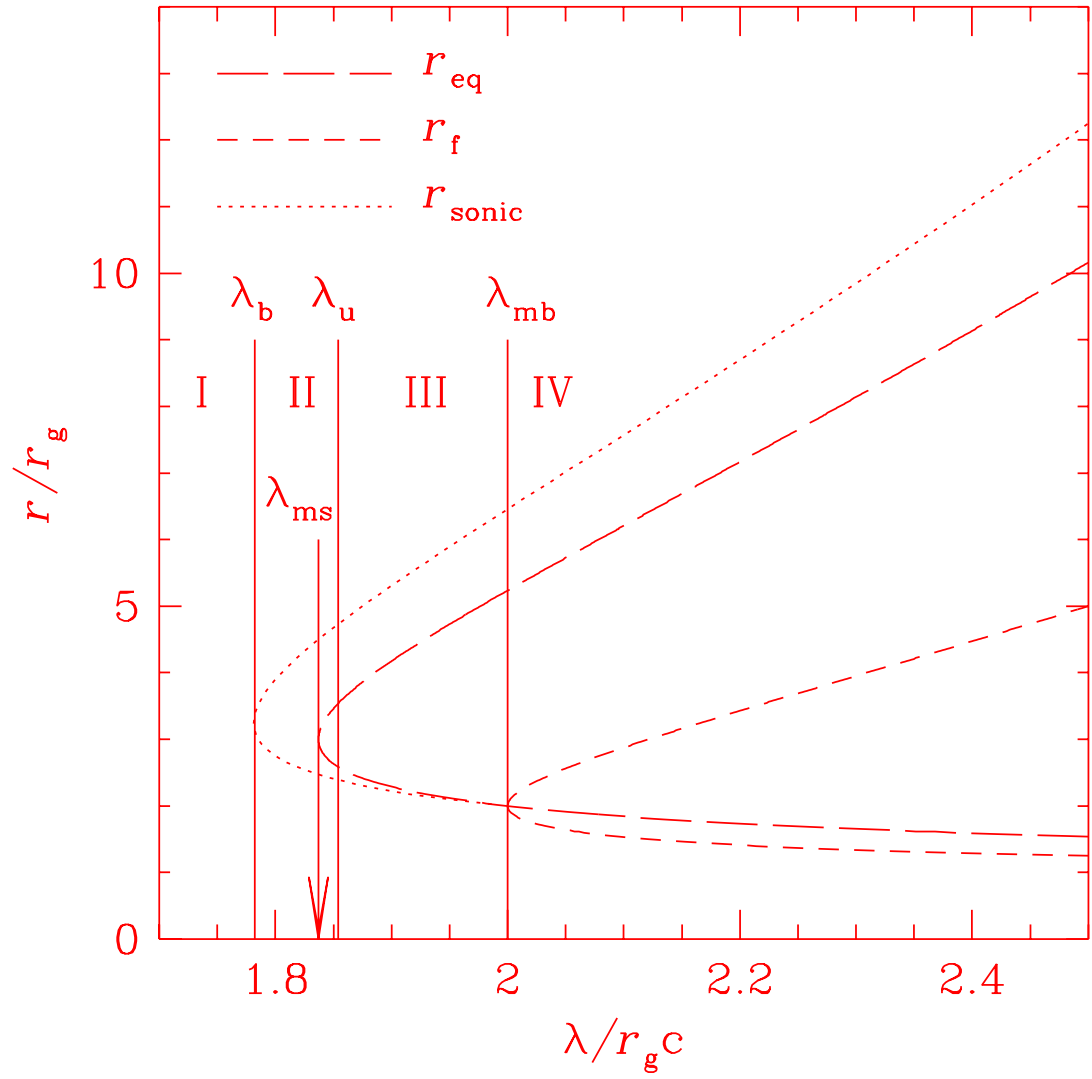
\item[Fig.~1] The radius of equal gravitational and centrifugal forces,
$r_{eq}$, the radius of the funnel wall along the equatorial plane,
$r_f$, and the locations of the sonic points of one-dimensional flows,
$r_{sonic}$, as a function of specific angular momentum.
The angular momentum space is divided into four Zones according to
whether sonic points and shocks exist or not.
Also, the locations of important angular momenta are marked.

\item[Fig.~2] The temporal evolution of the position of the accretion
shock along the equatorial plane, $r_{sh}$, the mass accretion rate
into the black hole, ${\dot M}_{acc}$, and the mean density, ${\bar\rho}$,
and the mean thermal energy, ${\bar e}_{th}$, of disk matter inside the
computational domain.
(a) is for Model M2 belonging to Zone I (left panel) and Model M5
belonging to Zone III (right panel).
(b) and (c) are for Models M3 and M4, respectively, belonging to Zone II.
The dotted curves in the mean density in (b) and (c) represent the
analytic fittings with Eqs.~(7) to (10).
Vertical dashed lines in (c) are drawn at the four epochs where the
panels of Fig.~4 are plotted.

\item[Fig.~3] The density contours and the velocity vectors at
$t=2\times10^4$ in Model M2 (left panel) and Model M5 (right panel).
The plots show the region with $0\leq r\leq36$ and $0\leq z\leq36$.
The density contours are drawn with $\rho=10^k$ with $k=-3$, $-2.9$,
$-2.8$, $\ldots$, $2$.
The velocity vectors are drawn at every fourth grid along  the $r$ and
$z$-directions.

\item[Fig.~4] The density contours and the velocity vectors at
$t=1.4\times10^4$, $1.6\times10^4$, $1.8\times10^4$, and $2\times10^4$
in Model M4.
The description on the plotted region and quantities is identical
to that of Fig.~3.

\item[Fig.~5] The density contours and the velocity vectors at
$t=2\times10^4$ in Model M7 (left panel) and Model M9 (right panel).
The plots show the region with $0\leq r\leq50$ and $0\leq z\leq50$.
The description on the plotted quantities is identical to that of Fig.~3. 

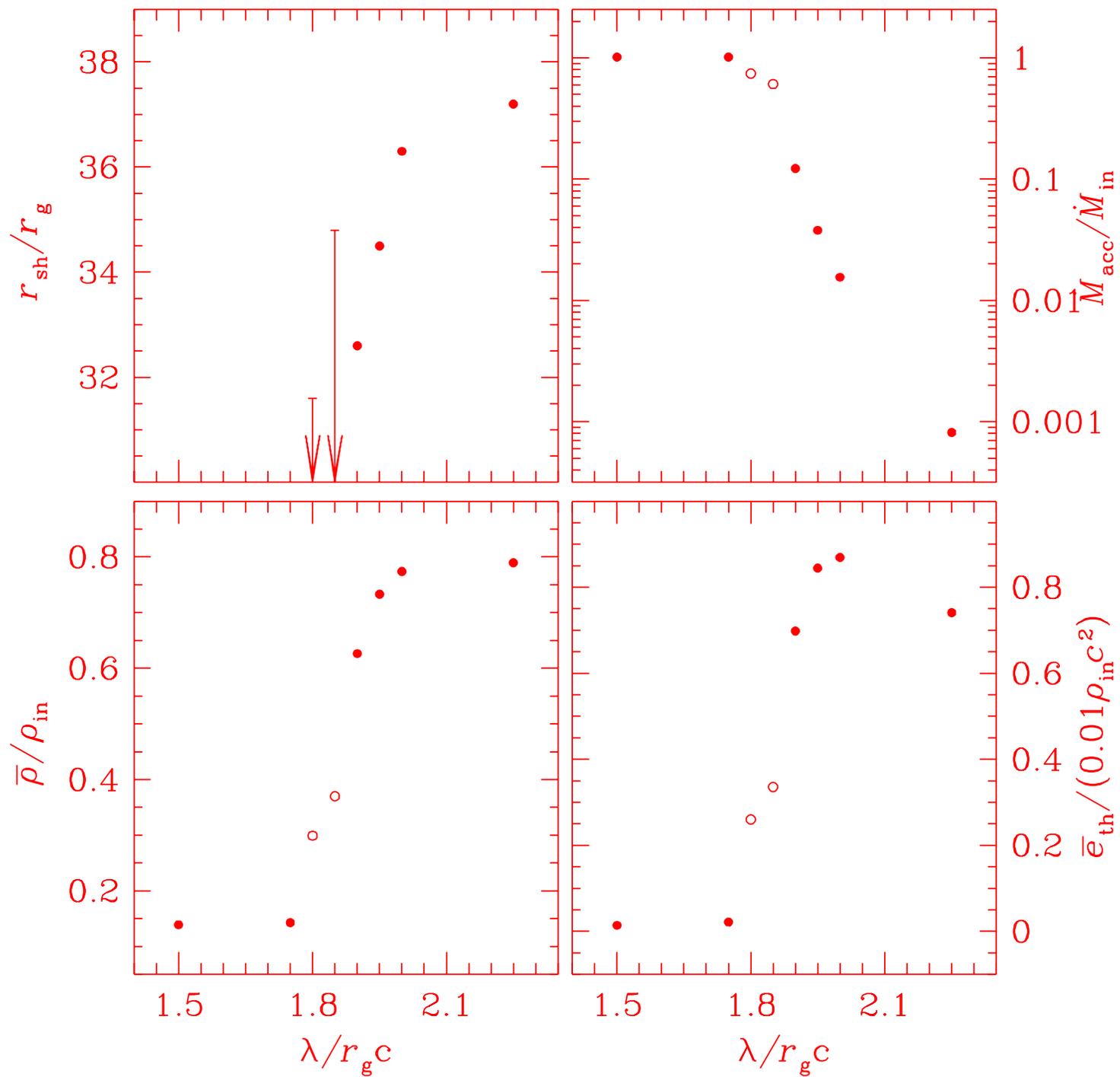
\item[Fig.~6] The time-averaged values of the position of the accretion
shock along the equatorial plane, $r_{sh}$, the mass accretion rate into
the black hole, ${\dot M}_{acc}$, the mean density, ${\bar\rho}$, and
the mean thermal energy, ${\bar e}_{th}$, of disk matter in Models M1 to M8.
In the cases showing a periodic behavior (Models M3 and M4), for the
position of the accretion shock the largest value is plotted with arrows
and for other quantities the averaged values over exact periods are
plotted with open circles.

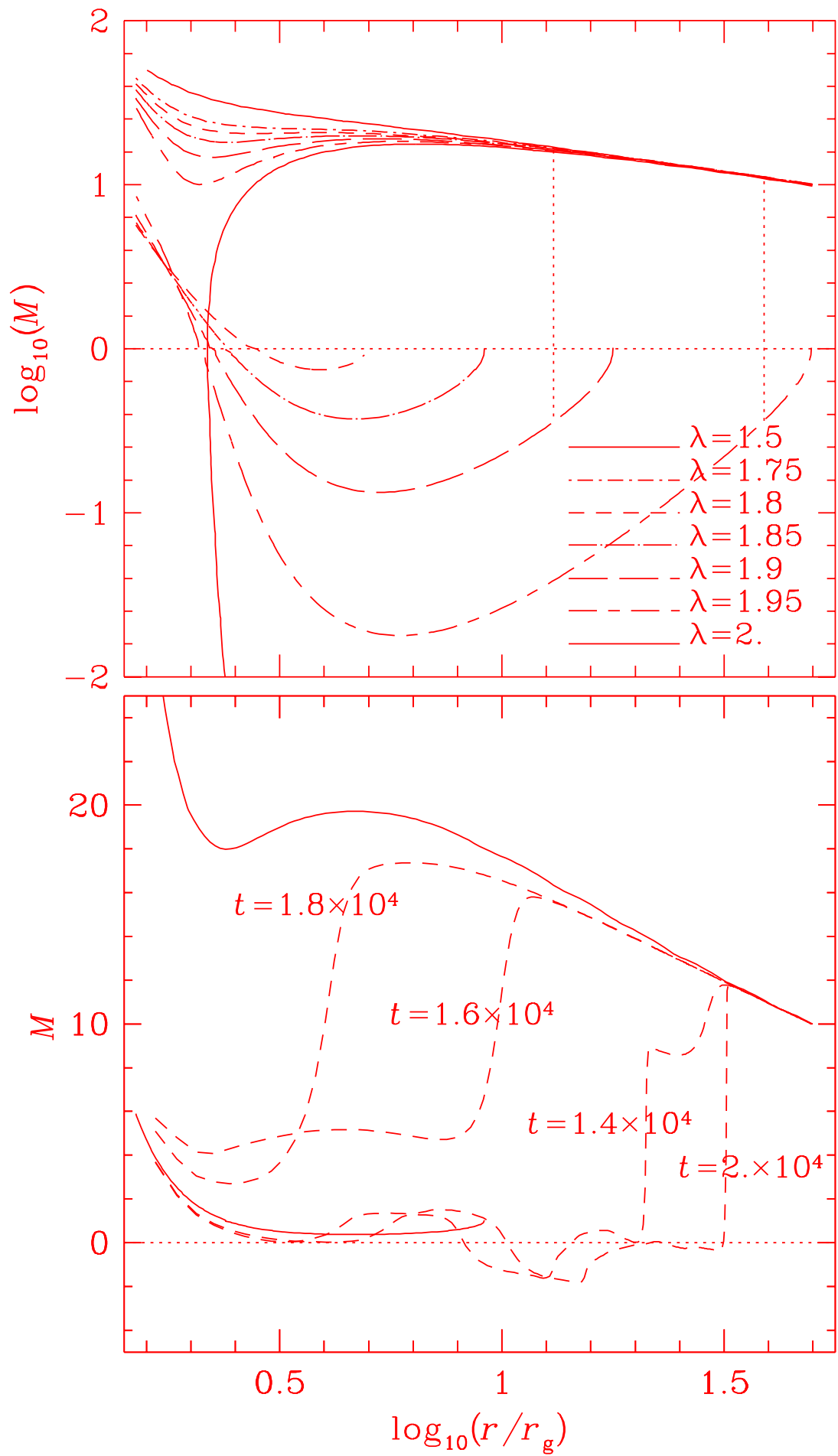
\item[Fig.~7] In the upper panel, the analytical solutions, Mach number
vs.~radial distance, of one-dimensional, zero-energy flows with $M=10$
at $r=50$ whose angular momentum ranges from $\lambda=1.5$ to $2$ are
shown.
Also the positions of the stable accretion shocks are indicated with
vertical dotted lines, in the cases that it exists (for $\lambda=1.9$
and $1.95$).
In the lower panel, the analytical Mach number distributions for
$\lambda=1.85$ (solid curves) are shown along with the numerical ones on
the equatorial plane (dashed curves) at four different phases in Model M4.

\end{description}

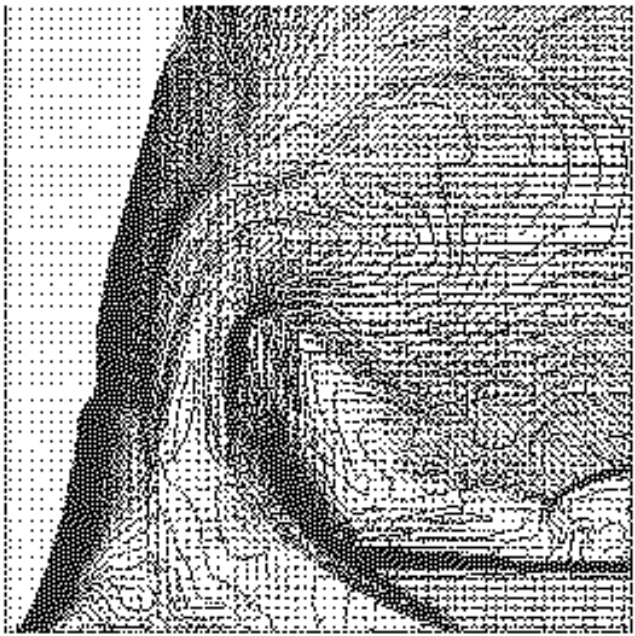
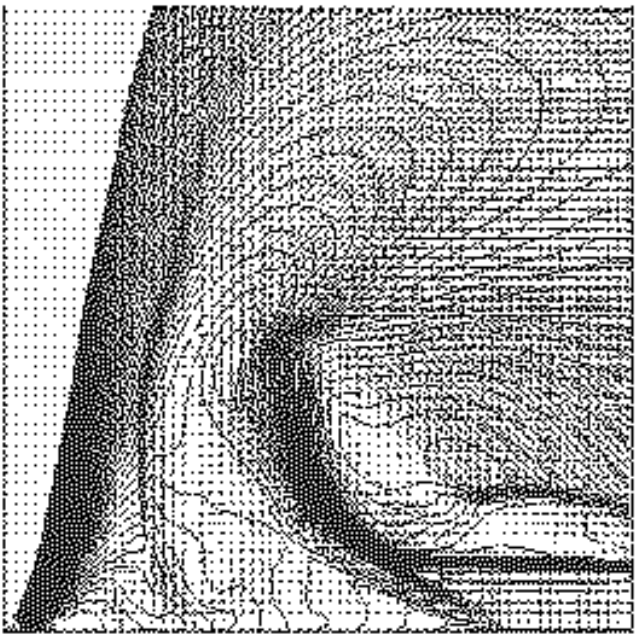
\end{document}